\documentclass[12pt]{JHEP3}

\usepackage{epsfig}
\epsfclipon

\def\hf{{1\over 2}}
\def\Dsl{\hbox{/\kern-.6700em\it D}} 
\def\dsl{\hbox{/\kern-.5300em$\partial$}}

\def\sla#1{\hbox{/\kern-.6700em #1}}

\def\eq{\begin{equation}}
\def\eeq{\end{equation}}
\def\eqa{\begin{eqnarray}}
\def\eeqa{\end{eqnarray}}

\def\bd{\begin{displaymath}}
\def\ed{\end{diplaymath}}

\def\Box{ {\,\lower 0.9pt\vbox{\hrule\hbox{\vrule height0.2cm \hskip 0.2cm \vrule height 0.2cm
}\hrule}\,}}
\def\lsim{{\ \lower-1.2pt\vbox{\hbox{\rlap{$<$}\lower5pt\vbox{\hbox{$\sim$}}}}\ }}
\def\gsim{{\ \lower-1.2pt\vbox{\hbox{\rlap{$>$}\lower5pt\vbox{\hbox{$\sim$}}}}\ }}

\def\pref#1{(\ref{#1})}

\def\hf{{1\over 2}}

\def\Dsl{\hbox{/\kern-.6700em\it D}} 
\def\Esl{\hbox{/\kern-.6700em\it E}} 
\def\dsl{\hbox{/\kern-.5300em$\partial$}}

\def\beginvector{\left( \begin{array}{c} }
\def\endvector{\end{array} \right)}
\def\endignore{}
\def\ignore#1\endignore{}

\def\Sph2{{\mathcal S}^2}

\title{Dimensionless Coupling of Bulk Scalars at the LHC}
\author{ P.-H. Beauchemin,${}^{1,2}$ G. Azuelos${}^{1,3}$and
C.P.~Burgess${}^2$
\\

${}^1$ Laboratoire Ren\'e J.-A. L\'evesque,
 Universit\'e de Montr\'eal, \\
 C.P. 6128, Montr\'eal, Qu\'ebec, Canada, H3C 3J7.\\

${}^2$ Physics Department, McGill University,
                3600 University Street,\\
                Montr\'eal, Qu\'ebec, Canada, H3A 2T8\\

${}^3$ TRIUMF,
4004 Wesbrook Mall, Vancouver BC, V6T 2A3\\
}

\abstract{ We identify the lowest-dimension interaction which is
possible between Standard Model brane fields and bulk scalars in 6
dimensions. The lowest-dimension interaction is unique and
involves a trilinear coupling between the Standard Model Higgs and
the bulk scalar. Because this interaction has a dimensionless
coupling, it depends only logarithmically on ultraviolet mass
scales and heavy physics need not decouple from it. We compute its
influence on Higgs physics at ATLAS and identify how large a
coupling can be detected at the LHC. Besides providing a
potentially interesting signal in Higgs searches, such couplings
provide a major observational constraint on 6D
large-extra-dimensional models with scalars in the bulk. }

\preprint{McGill-04/12}

\begin{document}

\section{Introduction}

One of the most surprising observations of recent years has been
the recognition that extra dimensions could be larger than a
micron across \cite{ADD} without being inconsistent with
observations. This picture is consistent with the relative
strength of gravity and the weak interactions provided that three
properties hold: ($i$) there are only two dimensions which are
this large (for a total of 6, including the 4 dimensions we see);
($ii$) the scale of gravity in the extra dimensions is of order a
TeV; and ($iii$) ordinary particles (apart from the graviton) are
trapped on a 3-brane, and so are unable to probe the existence of
the large dimensions.

Not surprisingly, this observation has spurred considerable
phenomenological study of the signatures for large extra
dimensions at machines like the Large Hadron Collider (LHC)
\cite{GRW,LEDColliderTh,invisible,BHatLHC,giota}. With a few
exceptions \cite{ABFLLM,HUGO} these studies have focussed
exclusively on the implications of gravitational physics in the
bulk, partly because this physics is the least
model-dependent.\footnote{Under the rubric `gravitational' we
include also discussions of 4D scalars and vectors which arise as
components of the higher-dimensional metric. The situation for
large extra dimensions is in sharp contrast with studies of extra
dimensions which are much smaller than micron scales, for which a
variety of phenomenological studies of potential LHC signals have
been investigated.} In broad terms these studies have concluded
that bulk gravitational physics can be detectable at accelerators
like the LHC if the extra-dimensional gravity scale, $M_g$, is of
order a few TeV. On the other hand, astrophysical bounds
\cite{LEDastrobounds,HR} typically require lower bounds like $M_g
\gsim 10$ TeV, making it not clear whether observational effects
should be expected at colliders.

Recently there have been several new proposals which suggest a
re-examination of the implications of these kinds of
large-extra-dimensional (LED) models for experiments, starting
from several different motivations. In particular, large extra
dimensions for which the bulk is supersymmetric are being studied
because of their potential implications for the cosmological
constant \cite{SLED,ADSWarped} as well as because of the novel,
non-MSSM realization of weak-scale supersymmetry breaking which
they imply \cite{MSLED}.

These models should predict a very rich phenomenology at the TeV
scale, which is only now beginning to be explored. Among the new
features which they imply is the existence of a multitude of new
states living in the bulk which are necessarily light and weakly
coupled because they are related to the graviton by supersymmetry,
and because supersymmetry is only broken in the bulk at energies
smaller than 1 eV. Furthermore, these models have plausible
ultraviolet completions (like string theory) which imply numerous
new states at TeV energies, such as massive string modes, Kaluza
Klein or winding modes associated with additional TeV-scale extra
dimensions. Indeed the existence of these new states (many of
which are expected to be much lighter than $M_g$) may allow the
astrophysical bounds on $M_g$ to be reconciled with the existence
of observable effects at colliders \cite{MSLED}.

In this paper we explore a different way in which LED models might
produce observable effects at the LHC, despite having the gravity
scale at 10 TeV or higher. This new mechanism relies on the
existence on our brane of the ordinary Higgs doublet (as would be
the case if the brane physics includes the Standard Model), and on
the existence of a fundamental scalar in the 6D bulk which is a
singlet under any bulk gauge transformations, and is not an exact
Goldstone boson (and so does not have a shift symmetry, of the
form $\phi \to \phi + \epsilon + \dots$).

If these two kinds of scalars exist, then gauge invariance and
general covariance allow them to couple to the brane scalars
through a trilinear interaction whose effective coupling, $a$, is
{\it dimensionless} and so is unsuppressed by inverse powers of
the extra-dimensional gravity scale, $M_g$. Because we have ruled
out symmetries which forbid such a coupling, it is likely to be
generated by loops even if it is excluded at the classical level.
We shall see that this makes the coupling plausibly of a size
which could be observed at the LHC, regardless of the size of
$M_g$.

We calculate the sensitivity to such a coupling at the LHC, and find
that searches for the Higgs using its decay into 2 photons can probe
effective couplings down to $a \gsim 0.09$. They can do so because of
the new process, $pp \to h \phi \to \gamma \gamma \Esl_T$, which this
new effective interaction opens up.\footnote{The signal of Higgs
production in association with missing energy to which we are led
resembles that of the 4-dimensional model ref.~\cite{BPtV}, which was
proposed as the most minimal model for dark matter.} The signal for
higgs production in association with bulk scalars turns out to be
fairly easy to detect because none of the Standard Model backgrounds
involve appreciable missing energy.

Our presentation is organized as follows. The next section
describes this lowest-dimension interaction, and summarizes its
properties. This is followed, in section 3, by a calculation of
the parton-level process of $h \phi$ production through gluon
fusion, as well as the ancillary calculations which are required
in order to promote the parton process into a cross section for
the hadronic reaction $pp \to h \phi$. Section 4 then presents the
results of detailed simulations of this process, modelling the
signals which would be seen within the ATLAS detector. In
particular, we describe in this section how the significance of
the signal depends on the various cuts which can be imposed. Our
conclusions are briefly summarized in Section 5.

\section{Lowest-Dimension Bulk Scalar Couplings}
In the class of models of present interest, all Standard Model
particles reside on a 3-brane embedded within a 6-dimensional bulk
space, whose gravity scale is $M_g \sim 10$ TeV. Indeed, it is
conservative to assume that the particle content on our brane
consists of no other fields besides those required by the Standard
Model \cite{MSLED}.

In general the bulk can be populated by a variety of degrees of
freedom, but within the context of supersymmetric large extra
dimensions (SLED) the bulk particle content would consist of the
states of one of the varieties of 6-dimensional supergravity. For
instance, the fields of 6D supergravity might consist of
\cite{6DSUSY1,6DSUSY2} the metric, $g_{MN}$, 2-form gauge
potential, $B_{MN}$, a dilaton (scalar), $\phi$, plus their
fermionic partners: the gravitino, $\psi_M$, and dilatino, $\chi$.
In addition there may also be 6D matter multiplets, such as might
be required by anomaly-cancellation for a chiral supergravity
\cite{6DAC}. These could involve gauge multiplets (containing a
gauge potential, $A_M$, and gaugino, $\lambda$), hyper multiplets
(involving scalars, $\varphi^i$, and fermions, $\omega^a$) or
others.

Although the bulk couplings of these fields are dictated by
supersymmetry, the absence of supersymmetry on the branes permits
the bulk-brane couplings to be more complicated. In general,
whatever these bulk-brane couplings are, the most important ones
for phenomenological purposes are usually those having the lowest
dimension, since these typically dominate the effective field
theory at energies, $E \ll M_g$.\footnote{We follow here the
spirit of ref.~\cite{Bigfit}, which gives a similar discussion of
the phenomenology of the lowest-dimension couplings of Standard
Model fields in 4 dimensions.} For this reason we focus our
attention in this paper on the lowest-dimension coupling which is
allowed between the generic bulk fields and the (purely
Standard-Model) fields on our brane.

For any model with bulk 6D scalars there is a unique interaction
which is possible between a bulk scalar and a brane-bound Standard
Model field for which the effective coupling involves a {\it
non-negative} power of mass. This is the effective trilinear
coupling between a general bulk scalar $\phi(x,y)$ (not
necessarily the dilaton), and the Standard Model Higgs doublet,
$H(x)$:
\begin{equation} \label{E:efflagrangian1}
   S_{int}  = - a \int d^4 x \; \sqrt{-g} \,
     H^\dagger H(x)\phi(x,y_b) \,,
\end{equation}
where $a$ is the dimensionless coupling. We use here coordinates
$x^\mu$, $\mu = 0,1,2,3$ to label dimensions parallel to the
Standard Model brane, and $y^m$, $m = 4,5$, for those dimensions
transverse to the brane. $y^m = y_b^m$ denotes the position of the
Standard Model brane within the bulk. All other effective
interactions necessarily have couplings which are suppressed by
inverse powers of $M_g$, assuming only that the bulk fields are
singlets under the Standard Model gauge group.

In principle, it is possible for an interaction like
eq.~\pref{E:efflagrangian1} to be forbidden by symmetries of the
bulk, such as by shift symmetries $\delta \phi = \epsilon + \dots$
which would occur if $\phi$ were a Goldstone boson. This kind of
symmetry is quite common in higher-dimensional supergravities, for
which the bulk scalars are often the coordinates of coset spaces,
$G/H$. Such a coupling can appear for the 6D dilaton, depending on
the microscopic origin of the dilaton and of the 3-brane in
question.\footnote{For instance, if the 6D dilaton is essentially
the 10D dilaton, then a coupling like eq.~\pref{E:efflagrangian1}
is present if the 3-brane arises as a D5 or D7 brane wrapped about
a 2- or 4- cycle in the dimensional reduction to 6 dimensions from
10. It does {\it not} arise classically if the 3-brane is a D3
brane or is an NS5-brane wrapped on a 2-cycle. In the mechanism of
cosmological constant suppression in the 6D SLED proposal there is
no dilaton-brane coupling at the classical level \cite{ADSWarped},
because of the classical scale invariance which this mechanism
presupposes.}

It must be borne in mind, though, that even should the coupling of
eq.~\pref{E:efflagrangian1} not be present classically, unless
precluded by symmetries it will be generated by loops and the size
of the resulting contribution to $a$ can be significant given the
non-decoupling of high mass scales and the strengths of the bounds
we shall find below. A representative size for such a coupling as
generated by one loop in 6 dimensions might then be set by
loop-counting factors, like $a \sim N/(4 \pi)^3$ with $N$ denoting
the number of particle species circulating in the loop. For the
simplest 6D supergravities \cite{6DSUSY1} $N$ is typically
$O(10-20)$, while for chiral 6D supergravities the requirements of
anomaly cancellation can imply $N \gsim O(1000)$. Depending on the
number of fields which contribute, an estimate for the
loop-induced coupling could well be $a \sim O(0.01 - 1)$. We
shall see that couplings at the upper end of this range might be
observable at the LHC.

In unitary gauge, we have $H = \left( { 0 \atop v + h(x)} \right)$
where $v$ is the expectation value which breaks the electroweak
gauge group, and in the absence of bulk-brane couplings $h(x)$ is
the physical scalar field. In terms of this, the coupling to the
bulk field becomes
\begin{equation} \label{E:efflagrangian2}
   S_{int}  = - a \int d^4 x \; \sqrt{-g} \,
     (v + h)^2 \, \phi(x,y_b) \,,
\end{equation}
We see that three separate kinds of couplings are implied between
the bulk scalar, $\phi$, and the Higgs scalar, $h$: ($a$) a linear
potential for $\phi$ of the form $a \, v^2 \, \phi$; ($b$) a
bulk-brane mixing term of the form $a\, v \, h \, \phi$; and ($c$)
a trilinear coupling of the form $a\, h^2 \, \phi$. We ignore the
first of these since it acts to shift the ground state of $\phi$
away from the value determined by the bulk scalar potential, and
none of our results depend on this value in any case.

The second type of interaction implies a mixing between $h$ and
the various Kaluza-Klein modes of the bulk scalar, $\phi$. This
mixing is removed by diagonalizing the resulting scalar mass
matrix, which for small $a$ leads to an $O(a)$ overlap between the
physical Higgs state and the bulk KK modes. The physics of this
mixing is very similar to previous discussions of gravi-scalar
mixing with brane modes, and leads to a non-negligible invisible
width~\cite{GRW,invisible} for the Higgs particle. We do not
further explore this width in detail, although its implications
would be important to understand once the Higgs is discovered and
its properties are being explored in detail.

In this paper we focus on the trilinear $h^2 \, \phi$ interaction,
which can play a role in the discovery signal of the Higgs. In the
next sections we compute the implications of this coupling for
Higgs production in association with missing energy.

\section{$h-\phi$ Production at Colliders}
\label{sect:processes}

The effective action
\begin{equation} \label{E:efflagrangian}
   S_{int}  = - a \int d^4 x \; \sqrt{-g} \,
     h^2 \, \phi(x,y_b) \,,
\end{equation}
can be used to compute the cross-section at tree-level for
production of bulk scalars, radiated by a Higgs boson in $p$-$p$
collisions at the LHC. Since Higgs production at the LHC is
dominated by gluon fusion, our interest at the parton level is
therefore in the reaction $gg \to h\phi$, where both final-state
particles are on shell. This process is similar to those studied
in ref.~\cite{HUGO}, although the effective couplings of $\phi$ to
fermions and gluons used in that reference have higher dimension,
and so are likely to be dominated by the coupling to the Higgs
used here (in models for which it is present).

For most values of the Higgs mass its dominant decays are
hadronic, but such decay channels can be easy to miss because of
the strong QCD background at the LHC. Consequently, we choose
instead to study the more rare, but cleaner, $h \to \gamma\gamma$
channel. The physical signal which bulk-scalar emission would
produce in this channel is then two photons plus missing energy,
as the scalar $\phi$ escapes into the extra dimensions.

\begin{figure}[thbp]
  \begin{center}
    \mbox{\epsfig{file=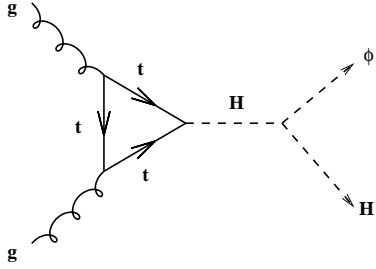, height=3.5cm}}
   \caption{Parton-level Feynman graph which dominates Higgs/bulk-scalar
   production at the LHC.}
    \label{fig:feyngraph}
  \end{center}
\end{figure}


We compute the Feynman rule for the $h-h-\phi$ vertex
and derive the parton-level cross-section
for producing a Higgs plus a bulk scalar in the final state. We
follow here the discussion of ref.~\cite{HUGO}, wherein the sum
over the finely-spaced Kaluza-Klein states of the $\phi$ field is
replaced by the equivalent integral over the extra-dimensional
phase-space of the bulk-scalar momentum. This, specialized to 2
large extra dimensions, gives the following phase-space measure
(after integration over the angular variable):
\begin{equation}
  \int_{\Omega_2}\frac{d^2L}{(2\pi)^2}=\frac{dM_\phi^2}{4\pi} \, .
\end{equation}
Here we denote the 6 components of the bulk-scalar momentum by $\{
\ell^\mu, L^m \}$, where $\ell^\mu$ denotes the components of
motion parallel to the brane and $L^m$ denotes its components in
the extra transverse dimensions. If $\phi$ has 6-dimensional mass
$\mu_\phi$ then $M_\phi^2 = \mu_\phi^2 + L^m L_m = - \ell^\mu
\ell_\mu$ is the effective 4-dimensional mass the bulk scalar
carries as seen by brane-bound observers due to its motion in the
extra dimensions. In later applications we take $\mu_\phi = 0$
because our interest lies with those scalars whose masses are kept
light, such as they would be if they were related to a massless
particle (like the graviton) by supersymmetry. Generalization of
our results to arbitrary $\mu_\phi$ is straightforward.

The dominant amplitude at the parton-level is obtained by
evaluating the Feynman graph of
Fig.~\pref{fig:feyngraph}.\footnote{Other graphs for which the
bulk scalar is emitted by other particles give contributions which
are suppressed relative to Fig.~\pref{fig:feyngraph} by powers of
external momenta divided by $M_g$.} In evaluating this graph we
use the full momentum-dependent expression for the effective
Higgs-gluon vertex which is obtained by evaluating the fermion
sub-loop, along the lines of refs.~\cite{HGGLoop}. It suffices to
keep only the contribution of the top-quark loop, in which case
the fermion loop leads to the following form factor for the
effective Higgs-gluon coupling (for on-shell gluons)
\begin{equation} \label{Hggeff}
   V_{ggh} = \frac{\alpha_s}{12 \pi v} \, {\cal F} \left(
   \frac{m_t^2}{Q^2} \right) \,,
\end{equation}
where $Q^2 = - q^2$, for $q^\mu$ the 4-momentum carried by the
virtual Higgs. We follow the conventions of ref.~\cite{NEWSCAL},
for which ${\cal F} (r) = 3[2r + r(4r-1) f(r)]$ and
\begin{equation} \label{fdef}
    f(r) = \left\{ \matrix{-2 \left[ \arcsin \left( {1 \over 2
    \sqrt{r}} \right) \right]^2 & \hbox{if}\quad r > \frac14 ; \cr \hf
    \, \left[ \ln \left( {\eta_+ \over \eta_-} \right) \right]^2 -
    {\pi^2 \over 2} + i \pi \ln \left( {\eta_+ \over \eta_-} \right) &
    \hbox{if}\quad r < \frac14 ; \cr} \right.
\end{equation}
with $\eta_\pm = \hf \pm \sqrt{\frac14 - r}$.

The form factor ${\cal F}(r)$ defined by the above expressions has
the following well-known properties \cite{HGGLoop}. It vanishes
for small $r$, ensuring the negligible contributions of all
low-mass quarks in the loop. For large $r$ it behaves as ${\cal
F}(r) = 1 + O(1/r)$. For intervening values of $r$, ${\cal F}(r)$
grows to a maximum for $r = O(1)$, falling off on either side
towards the two asymptotic limiting forms.\footnote{Because of the
small rise in ${\cal F}(r)$ for $m_t^2 < Q^2 < 4 \, m_t^2$, we
find that use of the asymptotic expression ${\cal F} \approx 1$
underestimates the size of the cross section by roughly 20\%.}

The differential parton-level cross-section obtained in this way
is given by:
\begin{eqnarray}
  \frac{d\sigma}{d\hat{t} \, dM_{\phi}^2}(gg \rightarrow h \phi) &=&
  \left( \frac{a^2 \alpha_{s}^2
    }{144 v^2 } \right) \,\frac{|{\cal F}|^2}{(\hat{s}- m_h^2)^2} .
\end{eqnarray}
where ${\cal F} = {\cal F}(m_t^2/\hat{s})$, $m_h$ is the Higgs
mass and $M_{\phi}$ is the effective bulk-scalar mass as defined
above. The quantities $\hat{s}$ and $\hat{t}$ denote the usual
parton-level Mandelstam kinematic variables. As usual, the
integrated cross section grows like $M_\phi^2$, and so is
dominated by the highest bulk-scalar masses, reflecting the
enormous extra-dimensional phase space which is available for such
states.

The cross-section for proton-proton collisions is obtained from
this parton-level result in the usual way, by convoluting with the
parton distribution functions, $f_i(x,Q^2)$. Since we are
interested in the production of a real Higgs that decays into two
photons for our later analysis, when performing these steps we
compute the cross-section for the missing-energy process $pp \to
\phi + h \to \gamma \gamma \Esl_T$. The result takes the following
form:
\begin{equation}
  \sigma(pp \to h \phi) =
  \int  dx_1 \, dx_2 \, d\hat{t} \, dM_\phi^2 \;
  \left[ f_g(x_1,Q^2) \, f_g(x_2,Q^2)
  \left( \frac{d\sigma(gg \to h \phi)}{d\hat{t}\,
  dM_\phi^2} \right)_{\hat{s}}
  \right]\,,
\end{equation}
where $\hat{s}$ is related to the $pp$ centre-of-mass energy,
$E_{cm}$, by $\hat{s} = x_1 \, x_2 \, E_{cm}^2$. Finally, we obtain
the cross section for $pp \to \gamma\gamma \Esl_T$ by multiplying
the above expression by the appropriate branching ratio, $B = B(h
\to \gamma\gamma)$.

We have calculated this cross section and implemented the process
in the generator PYTHIA~\cite{PYTHIA}. We generate events randomly
in phase space and assign weights to them. The events are then
accepted or rejected proportionately to these weights by PYTHIA,
which also performs the relevant hadronizations.

When performing the phase-space integrations we use the following
constraints:
\begin{itemize}
\item We require the transverse momentum of the final Higgs
particle to satisfy $P^2_T > P_{cut}^2$, where $P_{cut}$ is a
minimum value which can be chosen at generation level.
(We set $P_{cut}=0$ in the analysis below.)
This implies that the variable $\hat{t}$ must lie
in the range $t_{-} \le \hat{t} \le t_{+}$, with
\begin{eqnarray}
     t_{\pm} &=& \frac12 \Bigl[ {(m_h^2 + M_\phi^2-\hat{s}) \pm
     \sqrt{(m_h^2 +M^2_\phi - \hat{s})^2
    -4(m_h^2 M_\phi^2 + P_{cut}^2 \hat{s})}} \, \Bigr]
    \,. \nonumber\\.
\end{eqnarray}

\item Energy-momentum conservation implies the following upper
limit for $M^2_\phi$
\begin{equation}
            0 \leq M^2_\phi \leq M^2_{max} = \hat{s} +
            m_h^2 - 2\sqrt{\hat{s}(m_h^2 + P^2_{cut})} \, .
\end{equation}
\item The parton energy fractions, $x_i$, lie in the range
$x_{min} \le x \le 1$, with:
\begin{equation}
  x_{min} = \frac{\hat{s}_{min}}{s}
  = \frac{P^2_{cut}+\sqrt{P^2_{cut}+ m_h^2}} {s}
   \, ,
\end{equation}
and, as usual, $s$ denotes the Mandelstam initial-energy invariant
for the full proton-proton collision.
\end{itemize}

\begin{table}[htbp]
  \begin{center}
    \begin{tabular}{|c|c|c|c|c|c|c|c|c|}\hline
      Higgs mass (GeV) & 80 & 90 & 100 & 110 & 120 & 130 & 140 & 150 \\ \hline\hline
      Cross-section (pb) & 34.2 & 27.4 & 22.5 & 18.8 & 15.9 & 13.6 & 11.8 & 10.3 \\ \hline
      Branching ratio (\%) & 0.086 & 0.119 & 0.148 & 0.190 & 0.220 & 0.222 & 0.193 & 0.138 \\
\hline
      $\sigma \times B$ (fb) & 29.4 & 32.6 & 35.6 & 34.9 & 30.2 & 22.7 & 14.2 & 5.9\\ \hline
      SM $ pp \to h$ (pb) & 46.5 & 38.0 & 31.8 & 26.7 & 23.0 & 20.0 & 17.4 & 15.8\\ \hline
      Mass resolution (GeV) &  1.11 & 1.20 & 1.31 & 1.37 & 1.43 & 1.55 & 1.66 & 1.74\\ \hline
    \end{tabular}
    \caption{Cross-sections for the signal process $pp \to h \phi$
    using the coupling value $a=0.5$, and for $pp \to h X$ as a
    function of the Higgs mass.
    Also shown are the branching ratio for Higgs
            decay into two photons
        and the mass resolutions ($\sigma_H$)
        at high luminosity in ATLAS. }
    \label{tab:hgsccross}
  \end{center}
\end{table}

Table~\pref{tab:hgsccross} shows the various pieces of the total
cross section for $pp \to h \phi \to \gamma \gamma \Esl_T$ as a
function of Higgs mass in the range 80 GeV -- 150 GeV. The first
row of the table gives the total cross-section for $pp \to h \phi$
(in pb); the second row gives the branching ratio for the decay
channel $h \to \gamma\gamma$ (in percent), and the third row
multiplies these to give the cross section for $pp \to h\phi \to
\gamma\gamma \phi$ (in fb). For comparison, the last two rows give
the cross section for the Standard Model process $p p \to h X$, as
well as the Higgs mass resolution in the ATLAS detector, for the
process $h \to \gamma \gamma$, for high luminosity (as computed by
the ATLAS collaboration in ref.~\cite{TDR}). For this table,
the bulk-scalar process effective coupling constant, $a$, was set
to 0.5, so as to yield a $pp \to h\phi$ production cross section
which is comparable to the SM process $p p \to h X$. This gives a
rough indication of what size effective couplings might be
observable, and so motivates the more detailed calculations which
we now describe.

\section{More Detailed Simulations}

We now describe a more detailed analysis of the expected signals,
including Standard Model backgrounds in a more systematic way, as
well as incorporating detector effects. To do this we assign
parton flavors in each event according to the CTEQ 5L parton
distribution functions~\cite{CTEQ5L} evaluated at the renormalization scale
$Q^2=\frac12 ( {m_h^2}+ {M_\phi^2}) + p_T^2$  and colour flow
between these partons is applied. ATLAS detector effects were
incorporated using the fast Monte Carlo program ATLFAST
\cite{ATLFAST}.

\subsection{Standard Model Backgrounds}
Since the bulk scalar, $\phi$, radiated by the Higgs would quickly
escape into the extra dimensions, it should escape detection. The
observed process is therefore $p p \to \gamma\gamma$ with missing
energy in the final state, and so the backgrounds are the usual
ones for the process $pp \to h \to \gamma\gamma$. As discussed
in~\cite{TDR} these come in two types. First, there is an
irreducible background consisting of genuine photon pairs produced
by the Born process ($q\bar{q} \to \gamma\gamma$), by the box
diagram process ($gg \to \gamma\gamma$) and by quark
bremsstrahlung ($qg \to q\gamma \to q\gamma\gamma$). Second there
is also the reducible background --- consisting of QCD jet-jet or
$\gamma$-jet events --- in which one or both jets are
misidentified as photons~\cite{TDR}. These two sources of
background are comparable in size, even though the reducible
backgrounds have huge cross-sections compared to the irreducible
ones. This is because there are compensating large rejection
factors thanks to the efficient photon/jet discrimination which is
expected for ATLAS. These rejection factors have been evaluated to
be $2\times10^7$ or $8\times10^3$ respectively, for jet-jet and
$\gamma$-jet backgrounds. Once this rejection efficiency is
included, the reducible background events number about 20\% of the
expected number of irreducible background events.
  To the above backgrounds, we add processes with much lower cross
sections, but which include neutrinos in the final state. In particular,
we consider the associated production processes $Zh \to \nu\bar \nu \gamma\gamma$,
$Wh \to \ell\nu\gamma\gamma$ and $t\bar t h,~ h\to\gamma\gamma$.
We also take into account the processes $Z\gamma\gamma,~ Z\to \nu\bar\nu$
and $W\gamma\gamma,~ W\to\ell\nu$ which can also mimic the signal.
All the backgrounds were generated with PYTHIA. For the cases of $Z\gamma\gamma$ and
$W\gamma\gamma$, we simulated the processes $Z\gamma$ and $W\gamma$, with the
second $\gamma$ arising from initial or final state radiation. A $p_T$ cut of 35 GeV
was applied in these cases.

\begin{table}[htbp]
  \begin{center}
    \begin{tabular}{|c|c|c|}\hline
      Processes                                              & cross-section (pb)    & Number of events \\ \hline\hline
      $pp\to \gamma\gamma$ (Born)                            & 56.2                  & $5.62 \times 10^6$ \\ \hline
      $pp\to \gamma\gamma$ (box)                             & 49.0                  & $4.90 \times 10^6$ \\ \hline
      $pp\to$jet+jet                                         & $4.9 \times 10^8$     & $2.50 \times 10^6$ \\ \hline
      $pp\to$jet$+\gamma$                                    & $1.2 \times 10^5$     & $1.50 \times 10^6$ \\ \hline
      $pp \to h \to \gamma\gamma$                            & $4.63 \times 10^{-2}$ & 4630 \\ \hline
      $pp \to Zh,~Wh,~t\bar t h$                             &                       &      \\
       $Z \to \nu\bar \nu$, $W\to \ell\nu$, $h \to \gamma\gamma$  &$2.5\times 10^{-3}$ & 250  \\ \hline
      $pp \to Z \gamma      $; $Z \to \nu\bar \nu$           &   3.3                 & $3.3\times 10^5$   \\ \hline
      $pp \to W \gamma      $; $W \to \ell \nu$              &   5.6                 & $5.6\times 10^5$   \\ \hline

    \end{tabular}
    \caption{SM backgrounds to the production of bulk scalars in association
    with the Higgs at ATLAS,
         their cross-section (for an $E_T^{cut}$ of 23 GeV) and the total
     number of events expected at ATLAS for an integrated luminosity of
         100 fb$^{-1}$ (after application of rejection factors).}
    \label{tab:SMbackground}
  \end{center}
\end{table}

Table~\pref{tab:SMbackground} shows the cross-sections and the
total number of events expected after the application of the
above-mentioned rejection factors, for each background process.
The table assumes an integrated luminosity of 100 fb${}^{-1}$.
Notice that a further reduction factor of around 80\% must also be
applied in addition due to the expected reconstruction efficiency
of photons. Finally, following ref.~\cite{TDR}, we incorporate the
quark bremsstrahlung process into the simulation by scaling the
two other irreducible backgrounds by 50\% of the combined Born
plus box contribution, after having applied isolation cuts in
ATLFAST.

\subsection{Analysis}

We now describe a set of cuts which can be used to isolate those
$pp \to \gamma\gamma$ events which also involve significant
amounts of missing energy. We then use these cuts to quantify the
smallest size for $a$ which can be expected to be detectable.

We first establish our criteria for identifying two isolated
photons. In the ATLAS detector photons are detected if they are
emitted with pseudorapidity in the range $|\eta|< 2.5$. We
consider such photons to be isolated if their transverse momentum
satisfies  $P_T^{\gamma} > 5.0$ GeV, and if there is less than 10
GeV of energy deposited by all other particles within a cone of
radius $\Delta R = \sqrt{(\Delta\phi)^2 + (\Delta\eta)^2} < 0.4$
around the photon of interest.

Part of the reducible background consists of jets which are
misidentified as photons, and so we need also define our criteria
for jet reconstruction. For this we use the cone algorithm, with a
cone radius of $\Delta R = 0.4$, a pseudorapidity coverage of 5.0
and a minimal jet energy threshold of 10 GeV.

\begin{figure}[htbp]
  \begin{center}
    \mbox{\epsfig{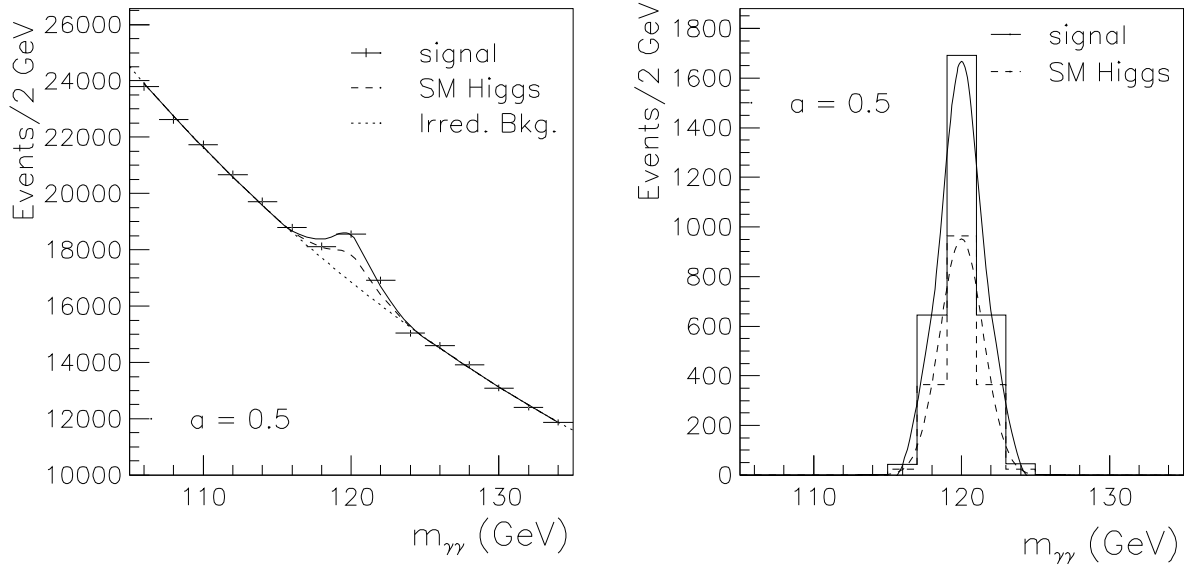}}
    \caption{Expected $h+\phi$ signal for $M_H=120.$ GeV, for an
    integrated luminosity of $100$ fb$^{-1}$ and with $hh\phi$ coupling
    value chosen as $a=0.5$. The left
         panel shows the signal on top of the irreducible
         background, while the right one shows the
         reconstruction of the mass peak.}
    \label{fig:hmass}
  \end{center}
\end{figure}

The first cuts to be imposed are those which optimize the
significance of the $h \to \gamma\gamma$ signal for the standard
Higgs search at ATLAS~\cite{TDR}. To this end we require:
\begin{quote}
{\bf Cut 1:} The two photon candidates, ordered in $p_T$, must
have transverse momenta which are in excess of 40 and 25 GeV.
(That is, we choose $P_T = 40$ GeV for photon 1 and $P_T =
25$ GeV for photon 2.)
\end{quote}

\begin{quote}
{\bf Cut 2:} Both photon candidates must lie in the pseudorapidity
interval $|\eta|<2.4$ and have a pseudorapidity separation of at
least 0.15 ($\Delta \eta>0.15$);
\end{quote}

\begin{quote}
{\bf Cut 3:} The reconstructed mass of the two photons, the two
jets, or the jet + $\gamma$ final state must have an invariant
mass which is sufficiently close to the Higgs mass.
Quantitatively, we demand: $M_H - 1.4\sigma_H < M_{\gamma\gamma} <
M_H + 1.4\sigma_H$, where $\sigma_H$ is the $h \to \gamma\gamma$
resolution quoted in table~\pref{tab:hgsccross}.
\end{quote}

Assuming an integrated luminosity of 100 fb${}^{-1}$, after
imposing these cuts and a 80\% efficiency for detecting each
photon, we are left with a total of 45,000 background events and
1,500 standard $h \to \gamma\gamma$ events. This leaves us with
38\% of the initial number of signal events --- $pp \to
\gamma\gamma + \Esl_T$ --- independently of $a$ (8930 events
before the cut for $a$ = 0.5).

In order to decide whether the missing-energy signal can be
winnowed out of the background, we first recall that the previous
ATLAS analyses \cite{TDR} indicate that these same cuts would
permit the standard Higgs signal to be identified with a
significance of 6.2 $\sigma$ (using $S/\sqrt{B}$ as the
significance criterion). After cuts, an effective coupling of size
$a = 0.5$ produces roughly the same number of $pp \to h \phi$
events as from the Standard Model $pp \to h$ process, for $m_h =$
120 GeV, even if the cross section is somewhat lower. This is due
to the fact that the final-state Higgs are more transverse in
energy, leading to a larger acceptance of photons. Therefore, we
roughly expect couplings of this size to be detectable at the 6
$\sigma$ level given 100 fb${}^{-1}$ of data. The significance for
the Higgs signal itself is thus doubled. Since the $h\phi$
production rate scales like $a^2$, a coupling $a = 0.44$ would
correspond to a 5 $\sigma$ significance. For couplings
this large about half of the Higgs particles are produced in
association with $\phi$ emission into the extra dimensions.

This situation is illustrated in Fig.~\pref{fig:hmass}, which
shows both the Standard Model and Higgs-$\phi$ production events
as a function of the invariant mass of the two photons, assuming a
Higgs mass of 120 GeV. The signal for Higgs-$\phi$ emission is
clearly visible on top of the irreducible background plus the
standard Higgs signal. These figures are also qualitatively the
same as those obtained for the standard $h \to \gamma\gamma$
process alone, which are published in the ATLAS Detector and
Physics Performance Report~\cite{TDR}.

\begin{figure}
  \begin{center}
    \mbox{\epsfig{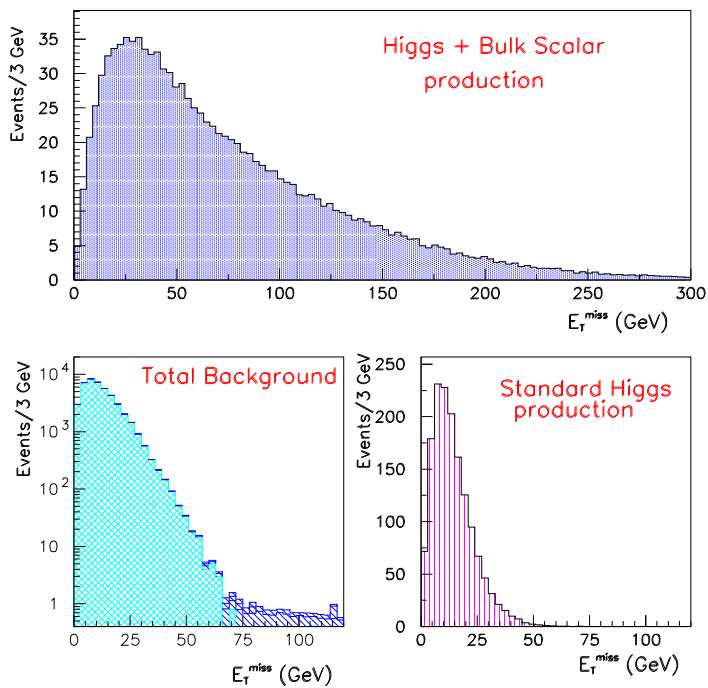}}
    \caption{Distribution of $\Esl_T$ for the Higgs + bulk scalar
    signal, assuming $a$ = 0.5 (top), the total background (bottom-left)
    and the $pp \to h$ process, with $m_H = 120$ GeV (bottom-right). The
    plots are normalized for an integrated luminosity of 100 fb$^{-1}$.}
    \label{fig:etmissdist}
  \end{center}
\end{figure}

As might be expected, and as we shall now see explicitly, those
events where Higgses are produced in association with bulk scalars
can be more efficiently identified by imposing a cut on the total
missing energy of the event. This is shown in
Fig.~\pref{fig:etmissdist}, which plots the number of background,
standard Higgs and $h\phi$ events as a function of the total
missing energy, $\Esl_T$. As this figure shows, very few
background or standard Higgs events have more than 50 GeV of
missing energy, while about half of the $h \phi$ events do.
The high energy tail in the background $\Esl_T$ distribution is
due principally to the processes $Zh$ and $Wh$.

\begin{figure}
  \begin{center}
    \mbox{\epsfig{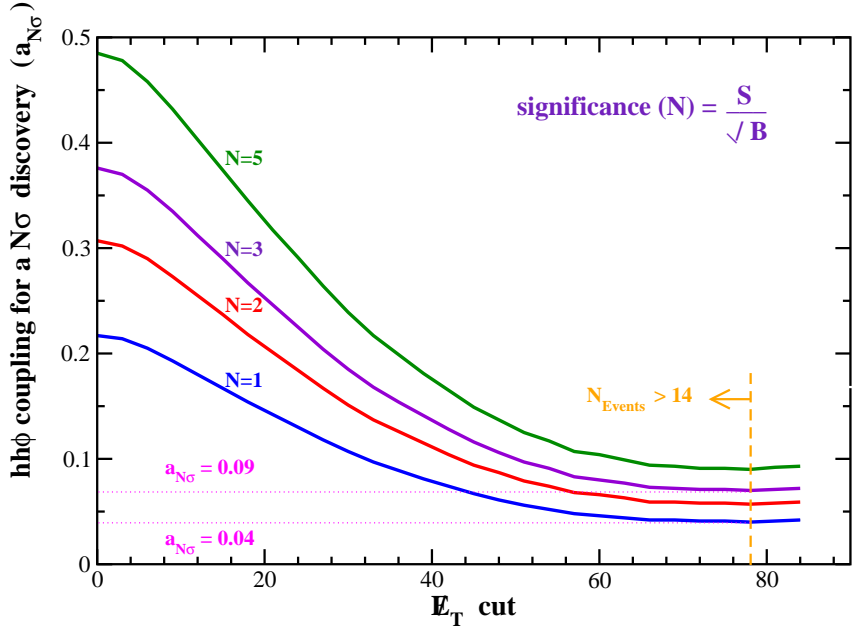}}
    \caption{Value of the $hh\phi$ coupling needed for
    different significances of the signal, as function of a
             cut on $\Esl_T$.}
    \label{fig:acoupl}
  \end{center}
\end{figure}

The larger the missing energy required in the event, the more the
background and standard Higgs events are excluded from the event
sample, but also the fewer $h\phi$ events there are.
Fig.~\pref{fig:acoupl} shows how this trade-off scales with the
effective coupling $a$, by showing the 5 $\sigma$ coupling reach
which is obtained as a function of the size of the missing energy
cut. In this figure the standard Higgs production is counted as
part of the background when computing the significance, since our
goal is to identify the $5\sigma$ discovery potential for the
particular process of Higgses produced in association with
$\phi$'s. If we define a discovery signal as a sample of at least
10 events which has significance greater than 5 $\sigma$, then the
smallest coupling for which discovery is possible (with 100
fb$^{-1}$ of data) is $a = 0.09$.

\begin{figure}
  \begin{center}
    \mbox{\epsfig{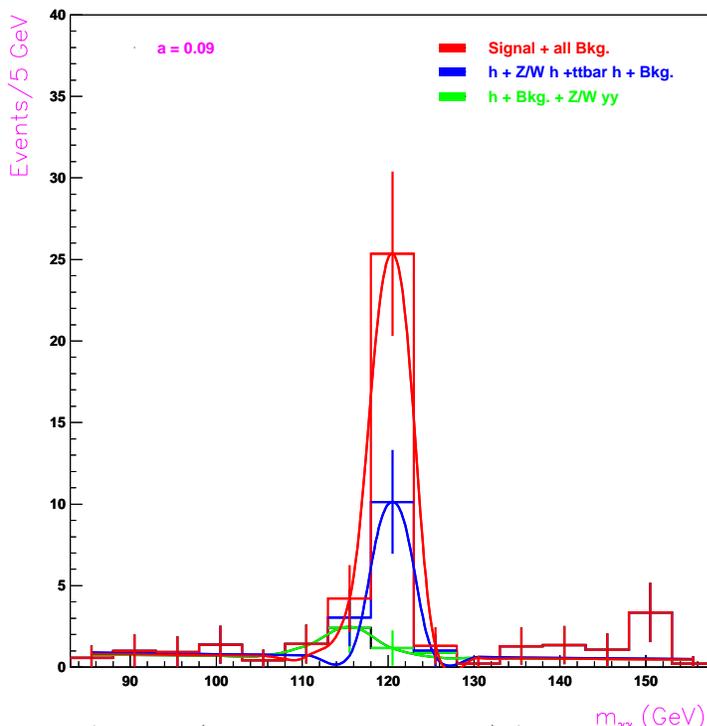}}
    \caption{Number of events (including backgrounds) for $h\phi$ and
    standard $h$ production, as a function of the two-photon invariant
    mass. This plot assumes the smallest-detectable coupling
      $a = 0.09$, and uses the optimal missing-energy cut,
      $\sla{E}_T > 78$ GeV.}
    \label{fig:acoupl1}
  \end{center}
\end{figure}

These considerations lead to the optimal missing-energy cut:
\begin{quote}
{\bf Cut 4:} The missing transverse energy of the entire event
must satisfy: $\Esl_T > 78$ GeV.
\end{quote}
Imposing such a cut, 14.3 signal events are left on a total
background of 8.2 events consisting of $\sim 0$ events of $\gamma\gamma$ + QCD,
8.0 events of $ h, Zh, Wh, t\bar t h$ and 0.2 events of $Z\gamma\gamma$
and $W\gamma\gamma$. Note that systematic errors on the measurement
of the $\Esl_T$ may be large. A proper evaluation of this uncertainty is
beyond the scope of this study but we do not expect that it will affect
significantly the main conclusions.
Fig.~\pref{fig:acoupl1} shows the number of
events vs invariant two-photon mass for the limiting case where $a
= 0.09$. We see from this figure that even this marginal case
yields a clear peak at the Higgs mass, leaving unambiguous
evidence for Higgs production in association with missing energy.

\begin{figure}
  \begin{center}
    \mbox{\epsfig{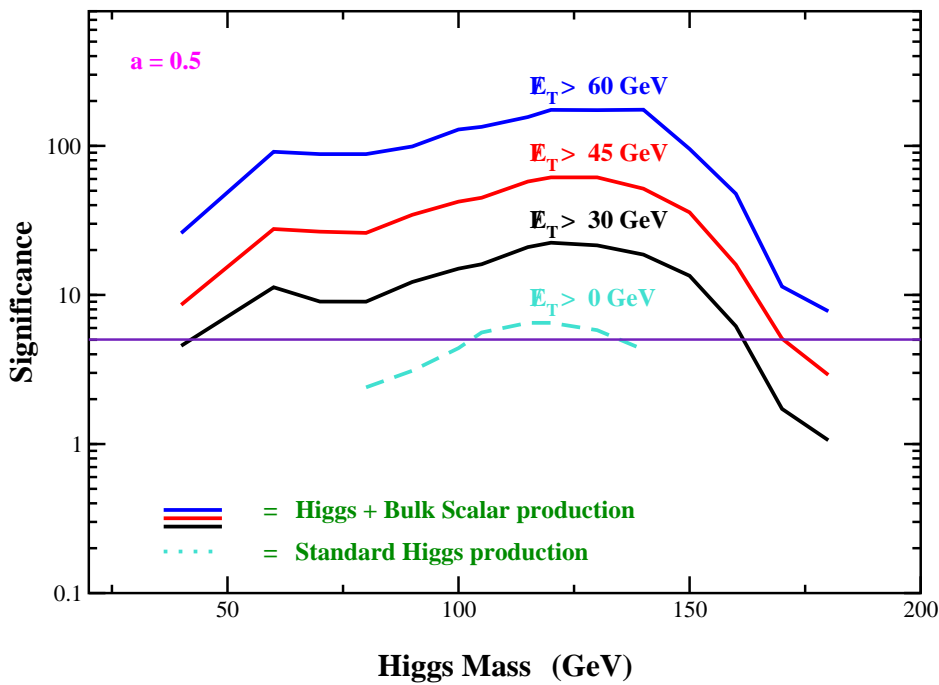}}
    \caption{The significance of the $\gamma\gamma$ signal alone as
    function of the Higgs mass for different
    values of the cut on $\Esl_T$. The figure assumes the choices
     $a=0.5$ and an integrated luminosity of 100 fb$^{-1}$.
     The dotted line corresponds to what can be obtained from the standard
     $h \to \gamma\gamma$ process making no cut on $\Esl_T$. This
     shows that a considerable gain in Higgs reach is possible should
     Higgs production be possible in association with missing energy.}
    \label{fig:signimass}
  \end{center}
\end{figure}

More generally, for larger values of $a$ than the above limit, the
significance for discovery of the Higgs boson itself can be much
improved, since the Standard Model backgrounds are considerably
reduced. This can be seen in Fig.~\pref{fig:signimass}, which
plots the significance of the $\gamma\gamma$ signal as a
function of Higgs mass, for several choices of missing energy cut.
As is clear from this figure, the curves with a nontrivial missing
energy cut are more significant than the one with no cut, simply
because of the dramatic reduction of background relative to signal
which the cut allows. The range of Higgs masses which are
accessible similarly increases, as can also be seen in
Fig.~\pref{fig:signimass}, by cutting on $\Esl_T$. For instance,
while the mass range accessible with no cut is $105 \;\hbox{GeV} <
m_h < 145 \;\hbox{GeV}$, this is extended to $60 \;\hbox{GeV} <
m_h < 180 \;\hbox{GeV}$ or more once cuts are applied. This figure
assumes $ a = 0.5$, but other values of the coupling are easily
incorporated using the result that the missing-energy cross
section scales as $a^2$.

We note in passing that the existence of a perturbative anomalous
coupling, eq.~\pref{E:efflagrangian1} should not invalidate the
earlier LEP searches for a Standard Model Higgs. On the other
hand, this coupling should enhance the number of events found when
searching with the Standard Model channel $e^+ e^- \to h Z \to
b\bar b \nu\bar\nu$, if a loose cut on the missing mass is
applied.

\section{Conclusions}

The analysis presented here reconsiders some of the observational
consequences of the existence of bulk scalars within a 6
dimensional scenario involving large extra dimensions. We have
done so motivated by the recent proposals of su\-per\-sym\-me\-tric large
extra dimensions, both as contributions towards understanding the
small size of the cosmological constant \cite{SLED} and as
alternative realizations of low-energy supersymmetry \cite{MSLED}.

We find that the lowest-dimension interaction which such a bulk
scalar can have with Standard Model fields on our brane has a
dimensionless coupling, $a$, and so can typically be expected to
be generated with a size which is {\it not} suppressed by inverse
powers of the 6D gravitational scale, $M_g \sim 10$ TeV. A
representative size for such a coupling as generated by one loop
in 6 dimensions might then be set by loop-counting factors like $a
\sim N/(4 \pi)^3 \sim 0.01-1$, given $N \sim 10-1000$ fields
circulating in the loop (which are plausible numbers for
supergravity theories in 6 dimensions). The effective coupling we
find resembles in some ways the effective brane-bulk mixing which
is possible for the bulk metric through an effective coupling of
the Higgs scalar to the curvature scalar on the brane
\cite{GRW,invisible}.

Using this effective interaction we compute the rate for the
process $pp \to h \phi \to \gamma \gamma \Esl_T$, in order to see
how large an effective coupling can be detected given reasonable
assumptions as to the performance of a detector like ATLAS at the
LHC. Our calculation assumes that the proton reaction is dominated
by the contribution of gluon fusion at the parton level.
Given the sensitivity to $a$
which we obtain, we believe there is sufficient motivation to go
back and perform more detailed studies of bulk-scalar production
at colliders. It must be noted that the coupling of a bulk scalar
to a more massive Higgs can lead to clean signatures, such as in
the case $h \to ZZ^{(*)} \to 4 \ell$.

We numerically integrate over the appropriate parton distributions
and include detector effects using existing ATLAS software. By
comparing the number of signal events to the expected Standard
Model backgrounds, we calculate the size of the effective
couplings to which experiments at the LHC can expect to be
sensitive. We find that couplings of order $a = 0.5$ imply that as
many Higgs particles are being produced in association with bulk
scalars as are being produced without them. We find that the
imposition of a missing energy cut $\Esl_T > 78$ GeV, greatly
improves the signal relative to background, and allows a 5
$\sigma$ detection of the effective interaction provided the
effective coupling is $a > 0.09$. These limits would begin to probe
the upper limit of the size of coupling which is obtained from a
generic 1-loop estimate.

We also notice that the existence of Higgs production in
association with missing energy is of considerable practical
interest in the detection of the Higgs itself. It allows
experiments to be sensitive to a much wider range of Higgs masses
(at a given level of significance) than would otherwise be
possible in the SM $\gamma\gamma$ decay channel.

We regard these results to be encouraging and --- together with
the strong motivation for bulk supersymmetry --- to further
motivate the study of the phenomenology of extra dimensional
fields (besides the higher-dimensional metric) within the
framework of large extra dimensions.

\section{Acknowledgments}
This work has been performed within the ATLAS collaboration. We
have made use of physics analysis and simulation tools which are
the result of collaboration-wide efforts. We thank Joaquim Matias
for helpful discussions. We also want to thank A. Parker and G. Unal
for useful comments. We would like to
acknowledge partial funding from NSERC (Canada). C.B.'s research
is also partially funded by FCAR (Qu\'ebec) and McGill University.

\end{document}